\documentstyle[prb,aps,twocolumn]{revtex}

\begin{document}
\draft
\twocolumn[\hsize\textwidth\columnwidth\hsize\csname @twocolumnfalse\endcsname

\title{Complicated nature of the gap in MgB$_{2}$: magnetic field-dependent optical
studies}
\author{H. J. Lee,$^{1}$ J. H. Jung,$^{1}$ K. W. Kim,$^{1}$ M. W. Kim,$^{1}$ T. W.
Noh,$^{1}$ Y. J. Wang,$^{2}$ W. N. Kang,$^{3}$ Eun-Mi Choi,$^{3}$ Hyeong-Jin
Kim,$^{3}$ and Sung-Ik Lee$^{3}$}
\address{$^{1}$School of Physics and Research Center for Oxide Electronics, Seoul\\
National University, Seoul 151-747, Korea}
\address{$^{2}$National High Magnetic Field Laboratory at Florida State University,\\
FL 32310, USA}
\address{$^{3}$National Creative Research Initiative Center for Superconductivity,\\
Department of Physics, Pohang University of Science and Technology, Pohang\\
790-784, Korea}
\date{\today }
\maketitle

\begin{abstract}
We investigated the magnetic field ($H$)-dependent optical conductivity
spectra of MgB$_{2}$ thin film in the far-infrared region. The $H$%
-dependences can be explained by the increase of normal metallic regions
embedded in the superconducting background. The area fraction of the normal
metallic region increases rather quickly at low field, but slowly at high
field. It follows neither $H$- nor $H^{1/2}$-dependences. The results
suggest the complicated nature of the superconducting gap in MgB$_{2}$.
\end{abstract}


\vskip1pc]
\newpage

The recent discovery of superconductivity in MgB$_{2}$ with $T_{C}=$ 39 K 
\cite{akimitsu} has generated a tremendous amount of attention among
condensed matter researchers. Many theoretical and experimental efforts have
been conducted to understand its superconductivity and the phonon-mediated
BCS mechanism\cite{budko} was suggested as being responsible. However, there
have been reports of distinctly different values of the superconducting gap $%
2\Delta $, i.e., from 3 to 16 meV, which makes understanding its physical
properties rather problematic. Also, the experimental evidences for the
deviation from an isotropic s-wave gap symmetry have been accumulating and
an anisotropic s-wave gap symmetry\cite{chen} or a multiple gap\cite{szabo}
have been suggested.

Optical measurements are known to be a powerful tool for investigating
important physical quantities, such as $2\Delta $, the scattering rate $%
1/\tau $, penetration depth $\lambda $, and plasma frequency\cite{IRers}.
Since the skin depth of light is about 1000 \r{A} in the far-infrared
region, optical measurements have an important advantage for obtaining bulk
properties compared to other surface sensitive techniques, such as tunneling
measurements. Also, optical measurements at high magnetic field ($H$) can
provide fruitful information, such as vortex dynamics, quasi-particle
excitation, and gap symmetry\cite{karr,choi}. There have been only a few
reports on the temperature-dependent optical properties of MgB$_{2}$\cite
{mgb2IR,jung,orenstein}. Earlier, we reported the value of $2\Delta
(0)\approx $ 5.2 meV and $2\Delta (0)/k_{B}T_{C}$ $\approx $ 1.8 for a $c$%
-axis oriented MgB$_{2}$ thin film \cite{jung}. Although our value of $%
2\Delta (0)/k_{B}T_{C}$ was half that of the BCS prediction, we found that $%
2\Delta $ seemed to follow the temperature dependences of the BCS formula.
To obtain further information on the gap nature of MgB$_{2}$, we have
performed $H$-dependent optical studies. [As far as we know, this is the
first investigation on the optical properties of MgB$_{2}$ under high $H$.]

In this paper, we report $H$-dependent complex optical conductivity spectra $%
\widetilde{\sigma }(\omega )$ of MgB$_{2}$ thin film. In the superconducting
state, the superconductivity became suppressed under the external $H$. The $%
\widetilde{\sigma }(\omega )$ of this mixed state can be modeled with the
Maxwell-Garnett theory, which assumes that the normal metallic regions are
embedded in a superconducting background. Using the composite medium theory,
the $H$-dependent $\widetilde{\sigma }(\omega )$ could be explained quite
well. Interestingly, the area fraction of the normal metallic region showed
neither an $H$ dependent characteristic of an $s$-wave superconductor (i.e.,
a linear dependence) nor that of a $d$-wave superconductor (i.e., $H^{1/2}$
dependence). It increased rapidly at low field, but rather slowly at high
field. This intriguing result suggests the existence of a complicated gap
nature for MgB$_{2}$.

We measured far-infrared transmission $T(\omega )$ and reflectivity $%
R(\omega )$ spectra of a MgB$_{2}$ thin film (thickness $\sim 200$ \r{A}) at
various $H$ from 0 to 17 T. A high quality $c$-axis oriented MgB$_{2}$ film
was grown on an Al$_{2}$O$_{3}$ substrate by the pulsed laser deposition
technique\cite{science}. The $dc$ resistivity measurement showed a sharp $%
T_{C}$ near 33 K. For optical measurements, the sample was cooled down to
4.2 K at zero field and then $H$ was applied along the $c$-axis. Using the
Bruker spectrophotometer, $T(\omega )$ and $R(\omega )$ were carefully
measured over the range 30 $\sim $ 200 cm$^{-1}$. For $T(\omega )$ mode, we
measured center area and for $R(\omega )$, whole area of the film.

Figure 1(a) shows the $H$-dependent $T(\omega )$ at 4.2 K. At 0 T, $T(\omega
)$ show a peak structure near 50 cm$^{-1}$, which is closely related to $%
2\Delta $\cite{jung}. With increasing $H$, the peak structure becomes
suppressed and the peak position moves to higher frequencies\cite{compare}.
At 17 T, the peak disappears and $T(\omega )$ show a flat response, i.e.,
normal metallic behavior. The corresponding $H$-dependent $R(\omega )$ are
shown in Fig. 1(b). At 0 T, $R(\omega )$ show a deep structure near 60 cm$%
^{-1}$. With increasing $H$, the deep structure in $R(\omega )$ becomes
broader and finally flattens at 17 T. It is interesting to note that
significant changes in $T(\omega )$ and $R(\omega )$ occur below 2 T\cite
{al2o3}.

We obtained $\widetilde{\sigma }(\omega )$ $(=\sigma _{1}(\omega )+i\sigma
_{2}(\omega ))$ from $T(\omega )$ and $R(\omega )$ in Fig. 1 by solving the
appropriate Fresnel equations for a film geometry\cite{ahn}. The light
reflects multiply both inside the film and the substrate. The multiple
reflections inside the film should be added coherently. However, because the
thickness of the substrate was much larger than the wavelength of the
incident light, the phase coherence can easily be lost due to surface
imperfections such as roughness, which will lead to the decrease of the
Fabry-P\'{e}rot fringes. To avoid such a problem, the optical measurements
were conducted with a low resolution (i.e., 4.0 cm$^{-1}$) and the multiple
reflections inside the substrate were added incoherently in the calculation.

Figure 2(a) shows the $H$-dependent $\sigma _{1}(\omega )$. At 0 T, $\sigma
_{1}(\omega )$ show a deep structure around 44 cm$^{-1}$ related with $%
2\Delta $ and a sharply increasing behavior above $2\Delta $. With
increasing $H$, the deep structure becomes smooth and $\sigma _{1}(\omega )$
at 17 T slowly decrease with increasing frequency. It should be noted that
the position of the deep structure increases as $H$ increases, which is
opposite to the temperature dependence of $2\Delta $\cite{jung}. Figure 2(b)
shows the $H$-dependent $\sigma _{2}(\omega )$. At 0 T, $\sigma _{2}(\omega
) $ show a $1/\omega $-dependence due to the delta-function of $\sigma
_{1}(0)$. $\sigma _{2}(\omega )$ at 4 T show a peak-like feature around 50 cm%
$^{-1}$ and $\sigma _{2}(\omega )$ at 17 T slowly increase with increasing
frequency. Note that the $\widetilde{\sigma }(\omega )$ at 0 T and 17 T are
similar to the reported behavior of the superconducting (i.e., 5 K) and the
normal metallic (i.e., 40 K) states, respectively\cite{orenstein}. From Fig.
2, we obtained $\sigma _{1}(0)\sim 20000$ $\Omega ^{-1}$cm$^{-1}$ and $%
1/\tau $ $\sim $ 800 cm$^{-1}$ in the normal state, and $2\Delta \sim $ 44 cm%
$^{-1}$ and $\lambda $ $\sim $ 2000 \r{A}\cite{pendepth} in the
superconducting state.

Such strong $H$-dependent $\sigma _{1}(\omega )$ and $\sigma _{2}(\omega )$
should be related to the evolution of a vortex in a type-II superconductor.
Below $H_{c1}$, the magnetic flux cannot penetrate into the superconductor,
which is the Meissner effect. Above $H_{c1}$, the magnetic flux starts to
penetrate into the superconductor, forming a vortex. Inside the vortex, the
superconducting regions become suppressed and turn into normal metallic
regions. With increasing $H$, the number of vortexes increases. Above $%
H_{c2} $, the superconducting properties are totally destroyed\cite{tinkham}%
. The values of $H_{c1}$ and $H_{c2}$ for MgB$_{2}$ were reported to be
around 450 Oe\cite{sharoni} and 20 T\cite{mhjung}, respectively.

Optical responses of a type-II superconductor under high $H$ have been
rarely investigated and there has not been much systematic analysis\cite
{choi}. To explain the optical responses shown in Figs. 2(a) and 2(b), we
used the composite medium theory, called the Maxwell Garnett theory (MGT).
In the long wavelength limit, the physical properties of the composite can
be described in terms of an effective dielectric constant, $\widetilde{%
\epsilon }^{eff}$. Since the vortexes are well isolated from each other due
to the inter-vortex Coulomb repulsion, as an approximation, we can consider
our film as a composite system composed of normal metal disks embedded in a
superconductor. Then, such a geometry can be approximated quite well by the
two-dimensional MGT \cite{twnoh}: $\widetilde{\epsilon }^{eff}(=4\pi i%
\widetilde{\sigma }^{eff}/\omega )$ can be written as,

\begin{equation}
\widetilde{\epsilon }^{eff}=\widetilde{\epsilon }^{s}\frac{(1-f_{n})%
\widetilde{\epsilon }^{s}+(1+f_{n})\widetilde{\epsilon }^{n}}{(1+f_{n})%
\widetilde{\epsilon }^{s}+(1-f_{n})\widetilde{\epsilon }^{n}},
\end{equation}
where $\widetilde{\epsilon }^{s}$ and $\widetilde{\epsilon }^{n}$ represent
the complex dielectric constants of the superconducting and normal metallic
regions, respectively, and $f_{n}$ represents the area fraction of the
metallic regions. For $\widetilde{\epsilon }^{s}$, we used the Zimmerman
formula\cite{zimmerman}, i.e., the optical response of the BCS
superconductor, with $2\Delta =44$ cm$^{-1}$ and $1/\tau =$ 800 cm$^{-1}$.
For $\widetilde{\epsilon }^{n}$, we used the simple Drude model with $\sigma
_{1}(0)=20000$ $\Omega ^{-1}$cm$^{-1}$ and $1/\tau =$ 800 cm$^{-1}$.

In Figs. 2(c) and 2(d), we show the calculated $\sigma _{1}^{eff}(\omega )$
and $\sigma _{2}^{eff}(\omega )$, respectively. It is clear that the
calculated $\sigma _{1}^{eff}(\omega )$ and $\sigma _{2}^{eff}(\omega )$ are
quite similar to the measured spectra. As shown in Figs. 1(a) and 1(b), the
calculated $T^{eff}(\omega )$ and $R^{eff}(\omega )$ from $\widetilde{\sigma 
}^{eff}(\omega )$ can also fit the experimental data quite well. These
results show that the MGT can describe the electrodynamic responses of a
type-II superconductor under high $H$ quite well.

The area fraction of the normal metallic regions for the MgB$_{2}$ film
shows quite an unusual $H$-dependent behavior. In Fig. 3(a), we plotted the $%
H$-dependent $f_{n}$, estimated by comparing the experimental data with the
calculated values, with error bars. $f_{n}$ increases sharply at low field
and slowly at high field; for example, more than half of the film becomes
metallic below $\sim $ 2 T $\ll H_{c2}$. It is known that the values of $%
f_{n}$ can be obtained from a heat capacity study by measuring the
coefficient $\gamma $. For an $s$-wave\ superconductor, $\gamma $ is known
to be proportional to $H$\cite{s-wave}. On the other hand, for a
superconductor with a node on the gap, for example, a $d$-wave symmetry, $%
\gamma $ should be proportional to $H^{1/2}$\cite{d-wave}. The dashed and
solid lines in Fig. 3 (a) show the $H$-dependent $f_{n}$ for $s$- and $d$%
-waves, respectively. Neither of these $H$-dependencies can explain our
experimental data. Fig. 3(b) shows the plot of log($f_{n}$) vs. log($H$).
[The error bars are smaller than the size of the symbols.] The two $H$
regions with different slopes are clearly seen. Interestingly, $f_{n}$ below
and above 2 T seems to follow the power law dependences of $H^{0.75}$ and $%
H^{0.33}$, respectively.

Recently, Bouquet {\it et al.}\cite{bouquet} observed a rapid increase of $%
\gamma $ at low $H$ and a saturation behavior at high $H$, which are in
qualitative agreement with our observed behavior of $f_{n}$. To explain this
behavior, they suggested the existence of two gaps. Szabo {\it et al.}\cite
{szabo} also reported the existence of two superconducting energy gaps using
point-contact spectroscopy measurements. Values of the small gap $2\Delta
_{S}$ and large gap $2\Delta _{L}$ were reported to be 5.6 meV and 14 meV,
respectively. Both superconducting gaps were shown to follow the temperature
dependence of the BCS formula. However, $2\Delta _{S}$ becomes strongly
suppressed with $H$ below 1.0 T.

There is some consistency between our observations and earlier works
proposing the two gap scenario\cite{liu}. Our observed value of $2\Delta
\approx $ 5.4 meV $(2\Delta /k_{B}T_{C}$ $\approx $ 1.9$)$ in the $ab$-plane
and its BCS temperature dependence seem to be consistent with the
characteristics of the small gap in the two gap scenario. Moreover, its
strong $H$-dependence at low field seems to agree with the results of Szabo 
{\it et al}. As shown in Fig. 3(b), the $H$-dependence of $f_{n}$\ seems to
have a crossover near 2 T. To explain this behavior in the two gap scenario,
we assumed that the reported value of $H_{c2}$, i.e., about 20 T,
corresponds to the large superconducting gap. With a crude approximation of $%
H_{c2}\sim 1/\xi ^{2}\sim \Delta ^{2}$, the corresponding $H_{c2}$ value for
the small superconducting gap, if it exists, will be around 2 T since $%
2\Delta _{S}\sim $ 5.4 meV. Then, the crossover behavior in Fig. 3(b) could
be explained in terms of a more rapid suppression of $2\Delta _{S}$ under $H$
in the two gap scenario\cite{comment}.

However, a simple two gap scenario based on two independent BCS-like
carriers with different gap values cannot explain our optical data. To
clarify this statement, we simulated $T(\omega )$ and $R(\omega )$ using the
two-fluid model, where the total optical conductivity spectra $\widetilde{%
\sigma }_{t}(\omega )$ can be written as

\begin{equation}
\widetilde{\sigma }_{t}(\omega )=f_{S}\widetilde{\sigma }_{S}(2\Delta
_{S},\omega )+(1-f_{S})\widetilde{\sigma }_{L}(2\Delta _{L},\omega ),
\end{equation}
where $\widetilde{\sigma }_{S}(2\Delta _{S},\omega )$ and $\widetilde{\sigma 
}_{L}(2\Delta _{L},\omega )$ represent the optical conductivity spectra with
the small and large gaps, respectively, and $f_{S}$ represents the fraction
of the superconducting carriers with $\Delta _{S}$. As shown in Figs. 4(a)
and 4(b), the experimental $T(\omega )$ and $R(\omega )$ at $H=0$ T can be
explained rather well only with $f_{S}\approx 1.0$. Note that all of the
earlier optical works on MgB$_{2}$ reports a small size gap only\cite
{mgb2IR,jung,orenstein}.

The apparent emergence of only the small gap feature in the optical spectra
as well as the unusual $H$-dependence of $f_{n}$ put new constraints on
understanding the MgB$_{2}$ superconducting gap in the two gap scenario\cite
{comment}. Although the $H$-dependence of $f_{n}$\ seems to be consistent
with the two gap picture, any two gap model does not explain why only the
small gap can be observed in the optical spectra of the $ab$-plane at $H=0$
T yet. The seemingly contradictory experimental facts suggest that the
nature of the superconducting gaps in MgB$_{2}$ should be related, which
will provide a complicated nature of the gap in the MgB$_{2}$. To clarify
the complicated gap nature of MgB$_{2}$, further studies are needed.

In summary, we have investigated the magnetic field-dependent optical
conductivity spectra of the MgB$_{2}$ thin film. The magnetic
field-dependent optical conductivity spectra could be explained by the
Maxwell Garnett theory, which assumes an increase of normal metallic regions
embedded in a superconducting background. The area fraction of the normal
metallic state was estimated and found to increase rather rapidly at low
field but slowly at high field. This magnetic field dependence provides a
new constraint to understand the multigap behavior of MgB$_{2}$. \ 

We thank H. K. Lee, Prof. J. Yu, and Y.-W. Son for useful discussions and
help in performing the experiments. This work at SNU and POSTECH was
supported by the Ministry of Science and Technology of Korea through the
Creative Research Initiative Program. This research was conducted at the
National High Magnetic Field Laboratory, which is supported by NSF
Cooperative Agreement No. DMR-0084173 and by the State of Florida.

\begin{figure}[tbp]
\caption{$H$-dependent (a) $\protect\sigma _{1}(\protect\omega )$ and (b) $%
\protect\sigma _{2}(\protect\omega )$. In (c) and (d), the calculated $%
\protect\sigma _{1}^{eff}(\protect\omega )$ and $\protect\sigma _{2}^{eff}(%
\protect\omega )$ using Maxwell Garnett Theory are shown, respectively.
Here, $f_{n}$ represents the area fraction of normal metallic region. }
\label{Fig:2}
\end{figure}

\begin{figure}[tbp]
\caption{(a) $H$-dependences of $f_{n}$. The dashed and solid lines
represent the power law dependences of $H$ and $H^{1/2}$, respectively. In
(b), we show the log-log plot of $f_{n}$ and $H$. The solid lines represent
the least-square fits. The exponents are 0.75 and 0.33 for below and above 2
T, respectively. }
\label{Fig:3}
\end{figure}

\begin{figure}[tbp]
\caption{(a) $T(\protect\omega )$ and (b) $R(\protect\omega )$ for various
fractions ($f_{S}$) of the small gap. The solid circles represent the
experimental data at 0 T. The solid, dashed, dot-dashed, dotted, and
dot-dot-dashed lines represent the corresponding spectra for $f_{S}=$ 0.0,
0.3, 0.7, 0.9, and 1.0, respectively. }
\label{Fig:4}
\end{figure}


\begin{references}
\bibitem{al2o3}  The optical constants of Al$_{2}$O$_{3}$ substrate are
nearly $H$-independent in the far-infrared region.
\bibitem{pendepth}  To evaluate superconducting penetration depth, we used $%
\lambda =(\mu _{0}\omega \sigma _{2}(\omega ,T))^{-1/2}$ with the value of $%
\sigma _{2}$ at lowest wavenumber. Here $\mu _{0}$ represents the magnetic
susceptibility at vacuum.
\bibitem{comment}  In the anisotropic s-wave scenario (Ref. 3), where the
large gap could be along the $c$-axis, the strong suppression of $f_{n}$
under low $H$ cannot be explained. 

\begin{figure}[tbp]
\caption{Experimental (symbols) and theoretical (solid lines) (a) $T(\protect%
\omega )$ and (b) $R(\protect\omega )$ for MgB$_{2}$ thin film at some
selected fields.}
\label{Fig:1}
\end{figure}
\end{references}
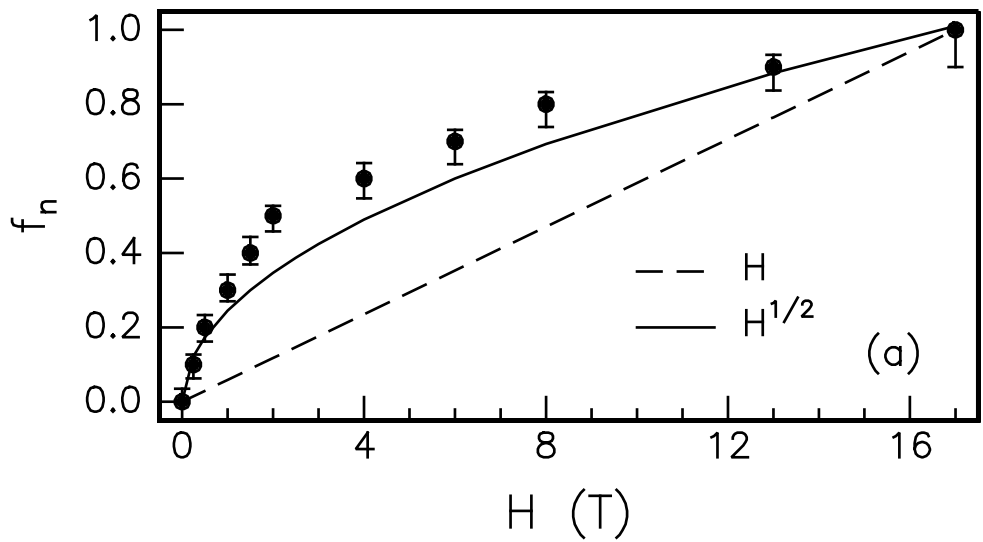
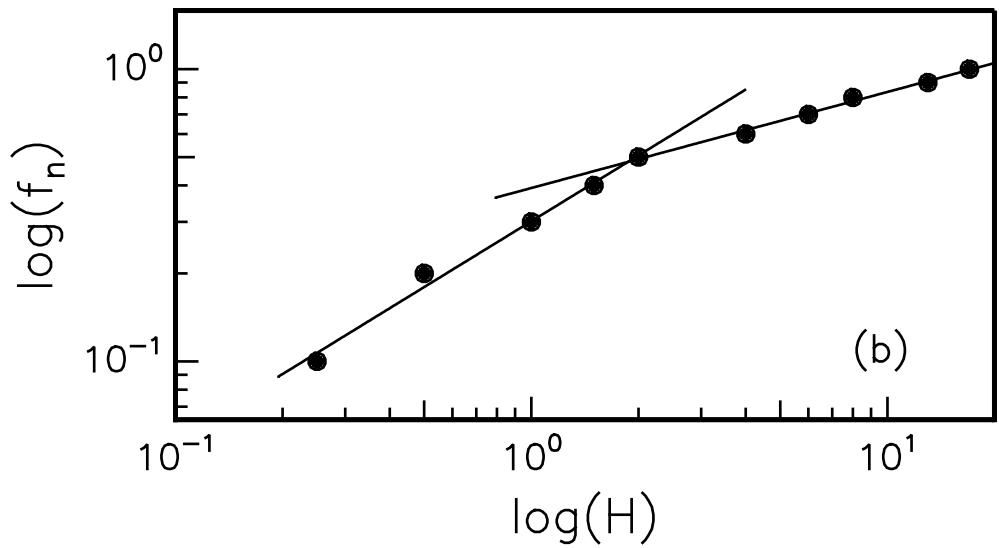
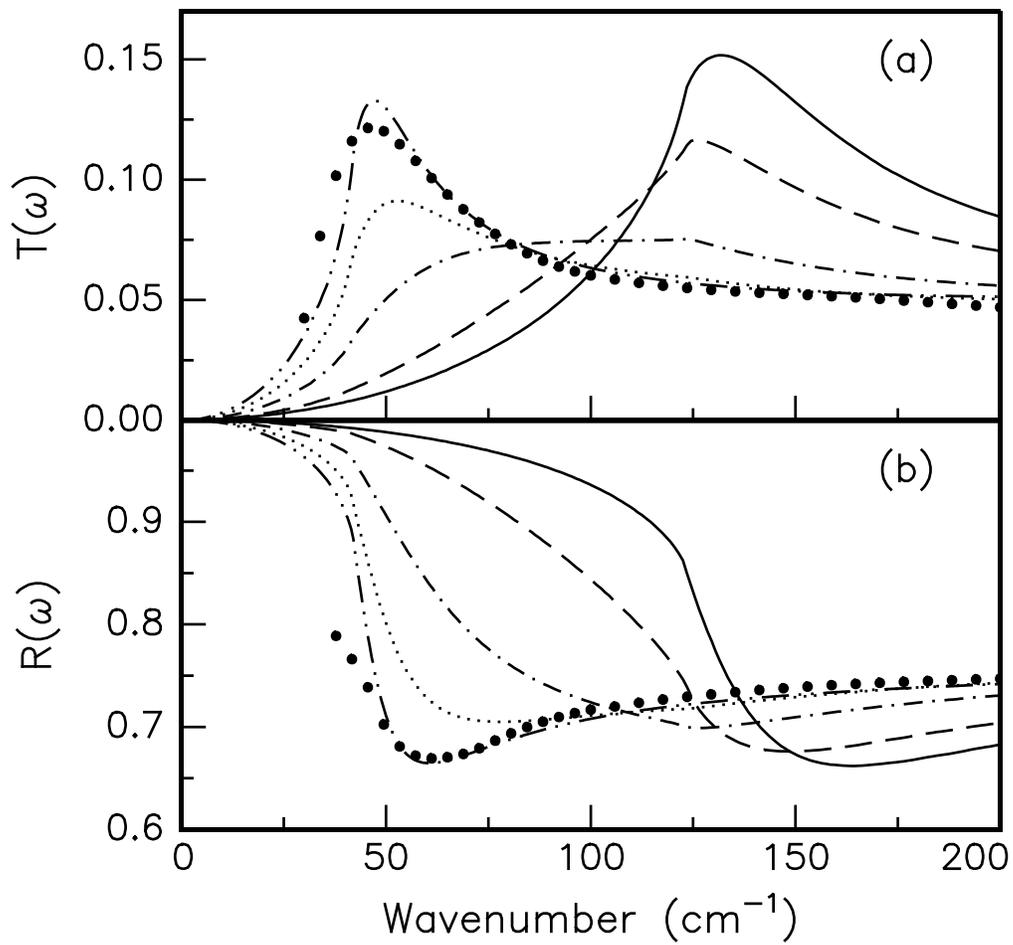
\end{document}